\PassOptionsToPackage{unicode}{hyperref}
\PassOptionsToPackage{hyphens}{url}
\PassOptionsToPackage{dvipsnames,svgnames,x11names}{xcolor}
\documentclass[
  british,
  11pt,
  letterpaper,
]{article}
\usepackage{xcolor}
\usepackage[margin=2.5cm]{geometry}
\usepackage{amsmath,amssymb}
\usepackage{iftex}
\ifPDFTeX
  \usepackage[T1]{fontenc}
  \usepackage[utf8]{inputenc}
  \usepackage{textcomp} % provide euro and other symbols
\else % if luatex or xetex
  \usepackage{unicode-math} % this also loads fontspec
  \defaultfontfeatures{Scale=MatchLowercase}
  \defaultfontfeatures[\rmfamily]{Ligatures=TeX,Scale=1}
\fi
\usepackage{lmodern}
\ifPDFTeX\else
\fi
\IfFileExists{upquote.sty}{\usepackage{upquote}}{}
\IfFileExists{microtype.sty}{% use microtype if available
  \usepackage[]{microtype}
  \UseMicrotypeSet[protrusion]{basicmath} % disable protrusion for tt fonts
}{}
\makeatletter
\@ifundefined{KOMAClassName}{% if non-KOMA class
  \IfFileExists{parskip.sty}{%
    \usepackage{parskip}
  }{% else
    \setlength{\parindent}{0pt}
    \setlength{\parskip}{6pt plus 2pt minus 1pt}}
}{% if KOMA class
  \KOMAoptions{parskip=half}}
\makeatother
\ifLuaTeX
\usepackage[bidi=basic,shorthands=off]{babel}
\else
\usepackage[bidi=default,shorthands=off]{babel}
\fi
\ifLuaTeX
  \usepackage{selnolig} % disable illegal ligatures
\fi
\providecommand{\tightlist}{%
  \setlength{\itemsep}{0pt}\setlength{\parskip}{0pt}}
\usepackage{graphicx}
\usepackage{microtype}
\usepackage{newunicodechar}
\newunicodechar{↔}{$\leftrightarrow$}
\usepackage{parskip}
\usepackage{booktabs}
\usepackage{array}
\usepackage{longtable}
\usepackage{caption}
\AtBeginDocument{\hypersetup{pdfcreator={LaTeX},pdfproducer={}}}
\usepackage[explicit]{titlesec}
\titleformat{\section}[block]{\normalfont\Large\bfseries}{\thesection.}{0.55em}{#1}
\titleformat{\subsection}[block]{\normalfont\large\bfseries}{\thesubsection}{0.55em}{#1}
\titleformat{\subsubsection}[block]{\normalfont\normalsize\bfseries}{\thesubsubsection}{0.5em}{#1}
\titlespacing*{\section}{0pt}{18pt plus 4pt minus 2pt}{6pt}
\titlespacing*{\subsection}{0pt}{12pt plus 3pt minus 2pt}{4pt}
\titlespacing*{\subsubsection}{0pt}{10pt plus 2pt minus 1pt}{3pt}
\usepackage{titling}
\pretitle{\begin{flushleft}\LARGE\bfseries}
\posttitle{\par\end{flushleft}\vskip 0.4em}
\preauthor{\begin{flushleft}\large}
\postauthor{\par\end{flushleft}\vskip 0.4em}
\predate{}
\postdate{}
\date{}
\usepackage{fancyhdr}
\usepackage{bookmark}
\IfFileExists{xurl.sty}{\usepackage{xurl}}{} % add URL line breaks if available
\makeatletter
\@ifundefined{xmpquote}{\newcommand{\xmpquote}[1]{#1}}{}
\makeatother
\hypersetup{
  pdftitle={Do Two On-Chain Observation Pipelines See the Same Tokens? Cross-Pipeline Coverage on the Solana pump.fun Launchpad},
  pdfauthor={Arati Uday Kamat},
  pdflang={en-GB},
  pdfsubject={Cross-pipeline overlap and coverage of two on-chain
observation pipelines on the Solana pump.fun launchpad},
  pdfkeywords={\xmpquote{Solana}, \xmpquote{pump.fun}, \xmpquote{decentralized
finance}, \xmpquote{memecoins}, \xmpquote{on-chain
measurement}, \xmpquote{data coverage}},
  colorlinks=true,
  linkcolor={NavyBlue},
  filecolor={Maroon},
  citecolor={NavyBlue},
  urlcolor={NavyBlue},
  pdfcreator={LaTeX via pandoc}}

\title{Do Two On-Chain Observation Pipelines See the Same Tokens?
Cross-Pipeline Coverage on the Solana pump.fun Launchpad}
\author{Arati Uday Kamat}
\date{}

\begin{document}
\maketitle

\textbf{Independent Researcher, United States.} ORCID:
\href{https://orcid.org/0009-0000-4781-312X}{0009-0000-4781-312X}. This
version supersedes version 1 of arXiv:2609.18975, the SSRN preprint
7128818 and Zenodo record
\href{https://doi.org/10.5281/zenodo.21399919}{10.5281/zenodo.21399919};
the corrections are summarised in Section 7.2.

\vspace{0.5em}

\section*{Abstract}\label{abstract}
\addcontentsline{toc}{section}{Abstract}

On-chain studies of memecoin launchpads usually rely on one
data-collection pipeline, yet whether differently configured pipelines
observe the same tokens is rarely measured. This paper compares the
output mint sets of two separately configured pipelines from one
research programme on the Solana pump.fun launchpad: a cohort-detection
pipeline that flags coordinated early-buyer wallets, and a
rejection-filtering pipeline that logs a trader-side observer's
pre-trade filter decisions. Two consecutive, non-overlapping windows are
analysed (v1: June 2026; v2: June-July 2026); in each, the rejection
data are restricted to the interval spanned by the cohort detections
under four timestamp rules. Only raw counts, observed proportions and
Jaccard/Dice indices are reported. In v2, where the rejection stream
covers the whole 17.5-day window, the cohort set contains 623 mints and
the rejection set 1,742, with 5 overlapping mints (0.803\% of cohort;
0.287\% of rejection). In v1 a collection gap means the rejection stream
covers only the final 14.25 hours (4.4\%) of the 13.4-day cohort window;
the 20,162 cohort and 53 rejection mints share 1 mint, and within the
covered interval none, a count too small to be informative. Both windows
show nearly disjoint outputs, agreeing in direction but not in
magnitude. Collector configuration, not only market behaviour, therefore
shapes which tokens an on-chain study observes, and the paper proposes
reporting cross-collector overlap and collector coverage as a routine
check. Data and a script that re-derives every number are openly
available.

\textbf{Keywords.} Solana; pump.fun; memecoins; on-chain measurement;
data coverage; decentralized finance.

\clearpage

\section{Introduction}\label{introduction}

On-chain studies of memecoin launchpads usually treat the output of one
data-collection pipeline as the market they describe. Studies of
pump.fun, the most prominent Solana launchpad, model which tokens
graduate from the bonding curve {[}1{]} or document manipulation at
scale {[}2{]}; detection studies of decentralised exchanges flag rug
pulls from transaction data {[}3, 4{]}; and earlier work on Ethereum and
BNB Smart Chain documents short-lived tokens, rug pulls and sniper bots
at token launch {[}5{]}. For cryptocurrency market data more generally,
the question of which data source to use has been examined directly
{[}6{]}. How far differently configured on-chain collectors observe the
same tokens has not, to the author's knowledge, been measured.

This paper measures that overlap for two separately configured pipelines
from one research programme on pump.fun: a \textbf{cohort-detection}
pipeline that flags persistent groups of wallets which buy early in the
same launches, and a \textbf{rejection-filtering} pipeline that logs the
pre-trade filter decisions of a trader-side observer. Both run on the
same on-chain event stream but subscribe to different slices of it,
apply different filters and target different stages of a token's life.
The question is descriptive: over the same window, do the two pipelines
see the same tokens?

The question has practical stakes for any analyst who characterises
participant activity on a permissionless launchpad from a single
observation pipeline. If the output mint sets of two collectors are
nearly disjoint, each collector sees only a fraction of the tokens that
the combined observation surface would show, and findings drawn from one
collector describe that collector's slice of the market rather than the
market. If the sets largely overlap, either collector gives a partly
redundant view of the same slice.

The answer is that the two pipelines see nearly disjoint token sets. In
the one fully covered window (v2, 17.5 days) they share 5 of 2,360
observed mints. In the earlier window (v1) a collection gap means the
rejection stream covers only the final 14.25 hours of the 13.4-day
cohort window, so v1 supports the direction of the finding but carries
much less weight (Section 5.3). The study does not estimate how much of
the pump.fun launch universe either pipeline covers. It shows that
collector configuration can determine which tokens an on-chain study
observes.

The paper makes two contributions.

\textbf{First}, a measurement of cross-pipeline overlap on pump.fun: raw
intersection counts, observed proportions on each side and
set-similarity indices for two consecutive windows, with the sensitivity
of the result to four timestamp rules for aligning the two streams.

\textbf{Second}, a simple check that on-chain measurement studies can
report: the overlap between collectors under a stated timestamp gate,
and each collector's own coverage of the analysis window. Section 5.3
shows why the second part matters: a single collection gap makes one of
the v1 proportions uninformative. The data and a standard-library script
that re-derives every reported number are openly available {[}7{]}.

The rest of the paper is organised as follows. Section 2 reviews related
literature. Section 3 describes the two pipelines and the datasets.
Section 4 states the window definitions, the timestamp rules and the
descriptive quantities reported. Section 5 presents the results,
including the rejection stream's coverage of each window. Section 6
discusses interpretation and implications for on-chain measurement.
Section 7 states the limitations and the corrections to an earlier
preprint, and Section 8 concludes.

\section{Related literature}\label{related-literature}

The strand closest to the observation mechanism measured here is
adversarial behaviour and value extraction on decentralised exchanges,
documented for Ethereum by Daian et al.~{[}8{]} and quantified on
Ethereum by Qin et al.~{[}9{]}. That literature measures extractable
value from public-mempool ordering and identifies wallet-level
extraction patterns; it does not compare the coverage of one observation
pipeline with another. Systematisations of decentralised-finance risk
are collected in Werner et al.~{[}10{]} and, for attacks specifically,
in Zhou et al.~{[}11{]}. Cross-chain MEV across nine EVM blockchains is
quantified in Öz et al.~{[}12{]}.

A second strand studies automated market makers. Empirical
characterisation of Uniswap as an automated market maker is provided by
Lehar and Parlour {[}13{]}. The theoretical properties of
constant-function market makers, including price-oracle behaviour and
optimal routing, are studied by Angeris and Chitra {[}14{]} and Angeris,
Evans, Chitra, and Boyd {[}15{]}. Reordering-manipulation risks and
their mitigations are systematised by Heimbach and Wattenhofer {[}16{]}.
The pump.fun setting inherits these AMM primitives after graduation but
differs in the pre-graduation phase, in which launches accumulate
liquidity along a deterministic bonding curve. Token adoption and
valuation on platforms are modelled by Cong, Li, and Wang {[}17, 18{]};
these models do not describe bonding-curve graduation, and this paper
does not import them.

A third strand addresses memecoin and launchpad markets. Empirical work
on meme-token trading includes Anton, Aptyka, Oesterreich, and Teuteberg
{[}19{]} on Dogecoin behavioural drivers and Conlon and Corbet {[}20{]}
on memecoin contagion and cryptocurrency risk. Cryptocurrency
pump-and-dump manipulation is characterised in Dhawan and Putniņš
{[}21{]} and in Tsuchiya {[}22{]}. On pump.fun specifically, Szwajcok et
al.~{[}2{]} analyse manipulation across the launchpad at scale, Marino
et al.~{[}1{]} model the probability that a token graduates from the
bonding curve, and Cordoba Otalora {[}23{]} presents an early student
research abstract; Solana token markets more broadly are analysed in an
SSRN working paper {[}24{]}.

A fourth strand concerns wallet clustering and the detection of harmful
tokens. The canonical treatment of address clustering is the Bitcoin
work of Meiklejohn et al.~{[}25, 26{]}; its logic, co-usage heuristics
on public-chain events, underlies both pipelines analysed here, although
this paper does not import a clustering estimator. Cernera et
al.~{[}5{]} study token ecosystems on Ethereum and BNB Smart Chain,
including rug pulls and sniper bots. Illicit-flow tracing and scam-token
detection are treated in Foley, Karlsen, and Putniņš {[}27{]} on illegal
activity financed through Bitcoin, in Xia et al.~{[}28{]} on Uniswap
scam tokens, in Lin, Chen, Wu, Zhang, Wang, and Zheng {[}29{]} on
smart-contract rug-pull warning, and in Srifa et al.~{[}3{]} and
Kalacheva et al.~{[}4{]} on rug-pull detection on decentralised
exchanges, the latter with a focus on meme coins. These studies each
rely on their own data collection; to the author's knowledge, none
compares the token sets that different collectors observe.

A fifth strand is the treatment of coverage when several partially
overlapping sources observe one population. Vidal-Tomás {[}6{]} examines
which cryptocurrency data sources scholars should use. Capture-recapture
methodology {[}30, 31{]} treats two or more observation sources as
capture events on a shared population and yields a lower-bound estimate
of that population plus a coverage rate for each source; record linkage
{[}32{]} treats the multi-source problem as a probabilistic-match
classification; and dual-frame survey inference {[}33{]} treats each
source as a frame and reweights to reduce coverage bias. These methods
need a plausible model of how units enter each source. In the on-chain
setting studied here, inclusion is deterministic given the collector
configuration (Section 4.3), which motivates the descriptive framing
adopted below rather than a formal capture-recapture estimator. An
epidemiological treatment of Berkson's bias and selection bias is
Westreich {[}34{]}; the framing here avoids inferential claims about the
pump.fun launch universe because the partial-observability mechanism is
not modelled.

Finally, on reproducibility, Peng {[}35{]} describes a spectrum of
standards for computational science, from publication alone to full
replication with linked code and data, and Menkveld et al.~{[}36{]}
show, in a study in which 164 research teams tested the same hypotheses
on the same data, that non-standard errors (variation in results across
teams) are sizeable relative to standard errors. The data and code
released with this paper aim at the full-replication end of Peng's
spectrum.

\section{Data and source pipelines}\label{data-and-source-pipelines}

\subsection{Cohort-detection pipeline}\label{cohort-detection-pipeline}

The cohort catalogue is the author's RED-COHORT-2026 dataset {[}37{]},
built with a two-stage detection pipeline. Stage 1 extracts the first
ten buyers of every qualifying launch. Stage 2 builds a co-occurrence
graph across launches, filters edges by weight, and surfaces persistent
wallet cohorts via union-find. Cohort membership at the mint level is
stable across the v1.0.0 through v1.1.1 releases of the dataset; the
incremental releases changed provenance notes only. The v1 catalogue
used here was captured during the June 2026 collection interval; the v2
catalogue was captured during the June-July 2026 interval by a different
version of the same detection pipeline. The v2 catalogue is not part of
the public RED-COHORT-2026 releases; it is published only in this
paper's data release {[}7{]}.

\textbf{Dependency and overlap disclosure.} The RED-COHORT-2026
catalogue (Zenodo concept DOI
\href{https://doi.org/10.5281/zenodo.20978741}{10.5281/zenodo.20978741})
is used here as a fixed input, not as a contribution. A companion
preprint {[}38{]} describes the detection approach and studies buyer
flow around coordinated cohorts using data from the same collection; the
present paper contributes only the cross-pipeline coverage measurement.

\subsection{Rejection-filtering
pipeline}\label{rejection-filtering-pipeline}

The rejection stream comes from a passive trader-side observer that
ingests candidate mints, scores them against a set of pre-trade filters,
and records a \texttt{rejectReason} for each mint that fails a filter.
The reasons recorded in the two files are price-trend filters (a 24-hour
downtrend and a multi-timeframe fade), a market-capitalisation range,
time-of-day filters, a fast-exit rule and, in v2 only, a
seller-dominance filter. Records include \texttt{mint},
\texttt{rejectTs}, \texttt{sampleTs}, and observation-time metadata. The
stream is predominantly, but not exclusively, a post-graduation
observation source: on the v2 window, 92.8\% of rejection rows carry
\texttt{dexId=pumpswap} and only 2.1\% carry \texttt{dexId=pumpfun}
(Section 3.5), and mints that never appear in a supported
post-graduation venue are unlikely to enter the stream at meaningful
scale.

\textbf{Relationship to the RED-REJECT-2026 dataset.} The two rejection
files are extracts of the stream published as the author's
RED-REJECT-2026-v2 dataset {[}39{]}, Zenodo concept DOI
\href{https://doi.org/10.5281/zenodo.21402476}{10.5281/zenodo.21402476}.
Every mint (83 in v1, 1,756 in v2) and every rejection event (312 and
6,823 unique mint and rejection-time pairs) in the two files also
appears in that dataset, so the files are not independent of it; the
present paper uses them only as the rejection-side mint sets. The
rejection stream contains no rejection events between 3 June 2026 (06:35
UTC) and 24 June 2026 (16:34 UTC), a collection gap that the
documentation of RED-REJECT-2026-v2 (version v2.2) does not mention;
that documentation describes the collection as continuous, and the
dataset's stated 96-day period includes the gap. Both statements are
re-derived here from that dataset's own file. The consequence of the gap
for the v1 window is reported in Section 5.3.

\textbf{Separately configured, not statistically independent.} The two
pipelines are separately configured observation streams within one
research programme. They differ in observation subscription, filter set,
target mint universe and observation time, and they share collector
infrastructure, licence and author. They are not claimed to be
statistically independent. This bears directly on the source-inclusion
assumptions of capture-recapture {[}30, 31{]} and dual-frame estimators
{[}33{]}, and it is one reason the analysis is descriptive rather than a
coverage-estimation exercise.

\subsection{Datasets}\label{datasets}

Four datasets underlie the analysis. Table 1 summarises them.

\begin{table}[htbp]
\caption{Datasets and observation windows underlying the analysis. Cohort windows run from the first to the last \texttt{detected\_at}; rejection windows from the first to the last \texttt{rejectTs} in each file. Bold-face figures denote the strict-window unique-mint counts used in Section 5.}
\centering\footnotesize
\resizebox{\linewidth}{!}{%
\begin{tabular}{@{}l l r r l l@{}}
\toprule
Dataset & Source & Rows & Unique mints & Window start (UTC) & Window end (UTC) \\
\midrule
Cohort v1            & RED-COHORT-2026 v1 & 20{,}163 & \textbf{20{,}162} & 2026-06-11T21:47:47Z & 2026-06-25T06:49:25Z \\
Cohort v2            & v2 detector run & 623      & \textbf{623}      & 2026-06-29T02:20:58Z & 2026-07-16T15:09:23Z \\
Rejection v1 window  & v1 rejection stream & 12{,}688 & 83 (file) / \textbf{53} (strict) & 2026-06-24T16:34:09Z & 2026-06-25T23:59:32Z \\
Rejection v2 window  & v2 rejection stream & 238{,}722 & 1{,}756 (file) / \textbf{1{,}742} (strict, Spec A) & 2026-06-29T00:02:11Z & 2026-07-16T15:47:36Z \\
\bottomrule
\end{tabular}%
}
\end{table}

No row in any of the four files is invalid JSON, every row carries a
\texttt{mint}, and every rejection row carries a \texttt{rejectTs}. The
rejection files mix two record layouts: 12,559 of the 12,688 v1 rows and
233,727 of the 238,722 v2 rows use the main layout, which includes
\texttt{sampleTs} and \texttt{dexId}; the remaining 129 and 4,995 rows
use an older layout with different field names (for example
\texttt{queriedAt}, \texttt{liquidityUsd}, \texttt{volumeH24}) and no
\texttt{sampleTs} or \texttt{dexId} (Supplementary Table S1). The four
files are pinned by SHA-256 checksums in the data release (Supplementary
Table S6).

\subsection{Mint notation}\label{mint-notation}

Specific mints are referred to by the first four and last four base-58
characters of their address (e.g., \texttt{5qjZ...pump}); the full
addresses appear in the supplementary material and in the released
mint-set files.

\subsection{Venue breakdown of the v2 rejection
stream}\label{venue-breakdown-of-the-v2-rejection-stream}

The v2 rejection file contains 238,722 rows across five \texttt{dexId}
values (Table 2).

\begin{table}[htbp]
\caption{Venue (\texttt{dexId}) breakdown of the v2 rejection stream.}
\centering\small
\begin{tabular}{@{}l r r r@{}}
\toprule
\texttt{dexId} & Rows & Row share & Unique mints \\
\midrule
pumpswap & 221{,}540 & 92.8\% & 1{,}666 \\
pumpfun  &   5{,}040 &  2.1\% &     41 \\
(none)   &   4{,}995 &  2.1\% & 1{,}612 \\
meteora  &   3{,}684 &  1.5\% &     41 \\
raydium  &   3{,}463 &  1.5\% &     29 \\
\bottomrule
\end{tabular}
\end{table}

The 5,040 rows / 41 distinct mints with \texttt{dexId=pumpfun} are a
residual: they were observed before or during graduation and reach the
rejection stream because the trader-side observer subscribes to the
pump.fun bonding-curve programme as well as to graduated DEX venues.
This residual is small, approximately 2.1\% of the v2 stream rows and
2.3\% of the file-level unique-mint count, but it tempers the
characterisation of the pipeline as predominantly post-graduation. The
``(none)'' rows are the older-layout records described in Section 3.3.

\section{Methods}\label{methods}

\subsection{Strict cohort window}\label{strict-cohort-window}

For each cohort file, define the strict cohort window as the closed
interval

\texttt{{[}min(detected\_at),\ max(detected\_at){]}}

over the mints in that file. This gate restricts the rejection dataset
to observations that occurred while the cohort detector was observing
candidate mints.

For v1, the strict cohort window is
\texttt{{[}2026-06-11T21:47:47Z,\ 2026-06-25T06:49:25Z{]}} (13.376
days). For v2, the strict cohort window is
\texttt{{[}2026-06-29T02:20:58Z,\ 2026-07-16T15:09:23Z{]}} (17.534
days).

The strict cohort window is set by the cohort detections alone; it does
not guarantee that the rejection stream was recording throughout the
window. Section 5.3 measures that coverage.

\subsection{Rejection-inclusion rule}\label{rejection-inclusion-rule}

Four timestamp rules for restricting the rejection stream to the strict
cohort window are considered:

\begin{itemize}
\tightlist
\item
  \textbf{Spec A}: \texttt{rejectTs} falls within the strict cohort
  window
\item
  \textbf{Spec B}: \texttt{sampleTs} falls within the strict cohort
  window
\item
  \textbf{Spec C}: either \texttt{rejectTs} or \texttt{sampleTs} falls
  within the strict cohort window (union rule)
\item
  \textbf{Spec D}: both \texttt{rejectTs} and \texttt{sampleTs} fall
  within the strict cohort window (intersection rule)
\end{itemize}

Under all four rules the unique-mint count and the overlap count are the
same on v1 (53 mints, 1 overlap) and differ only at the window
boundaries on v2 (1,742 or 1,749 mints, 5 overlap).

\subsection{Descriptive quantities reported; no sampling-model-based
CIs}\label{descriptive-quantities-reported-no-sampling-model-based-cis}

Let \emph{C} be the cohort mint set and \emph{R} be the strict-window
rejection mint set. The reported quantities are:

\begin{itemize}
\tightlist
\item
  Intersection count \(|C \cap R|\)
\item
  Union count \(|C \cup R|\)
\item
  Observed proportions \(100 \cdot |C \cap R| / |C|\) and
  \(100 \cdot |C \cap R| / |R|\)
\item
  Jaccard index \(|C \cap R| / |C \cup R|\) and Dice coefficient
  \(2 |C \cap R| / (|C| + |R|)\)
\end{itemize}

\textbf{Statistical framing.} The observation-selection process for each
pipeline is deterministic given the collector configuration, the Solana
Websocket subscription, the filter set and the collection window. An
observed mint is not a Bernoulli draw from an identifiable population;
there is no stochastic sampling unit on either side and no defensible
target population over which ``the true overlap rate'' would be defined.
This is the selection-bias setting reviewed by Westreich {[}34{]}: when
the selection mechanism is not modelled, inferential statements about
the underlying population are not supported. Confidence intervals for
the overlap proportions are therefore \textbf{not} reported, and no
statistical significance is claimed against any null. The reported
percentages are observed proportions of the observed sets, not estimates
of a population parameter.

All quantities are computed by a standard-library Python script that
reads the four input files, re-derives every reported number and stops
if any headline value differs from the value reported here {[}7{]}
(Supplementary Section S8).

\subsection{Sensitivity: wider extraction
window}\label{sensitivity-wider-extraction-window}

Both rejection files were extracted from the live stream using an
interval wider than the strict cohort window. On the v1 file, the
extraction window was
\texttt{{[}2026-06-11T21:47:47Z,\ 2026-06-26T00:00:00Z{]}}, 17h 10m 35s
wider than the strict cohort window. The additional rows increase the v1
unique-mint count from 53 (strict) to 83 (file) and the overlap count
from 1 to 2. These wider-window numbers are treated as a sensitivity
analysis only (Section 5.4). On v2 the difference between the strict and
file-level counts is small (1,742-1,749 strict vs 1,756 file) and has no
material effect on the observed proportions. Because of the collection
gap described in Section 3.2, the earliest row of the v1 file is at
2026-06-24T16:34:09Z, so every v1 rejection row lies in the last 31.4
hours of the extraction interval.

\section{Results}\label{results}

\subsection{Primary result: dual-window descriptive analysis (strict
cohort
window)}\label{primary-result-dual-window-descriptive-analysis-strict-cohort-window}

Table 3 reports the primary dual-window result under the strict cohort
window (Spec A, \texttt{rejectTs} within window). All four specs
(A/B/C/D) yield identical mint sets and overlap counts on v1; on v2 the
rejection mint count ranges from 1,742 (Specs A and D) to 1,749 (Specs B
and C), with the overlap count fixed at 5. The v1 row compares a
13.4-day cohort window with a rejection stream that covers only its
final 14.25 hours (Section 5.3).

\begin{table}[htbp]
\caption{Primary dual-window overlap results (strict cohort window; Spec A; observed proportions, no confidence intervals, see Section 4.3). Panel 3a reports overlap counts and observed proportions; Panel 3b reports set-similarity indices.}
\centering\small
\textit{Table 3a. Overlap counts and observed proportions.}\\[2pt]
\begin{tabular}{@{}l r r r r r r@{}}
\toprule
Window & Cohort mints & Rejection mints & Union mints & Overlap mints & \% of cohort & \% of rejection \\
\midrule
v1 & 20{,}162 &      53 & 20{,}214 & \textbf{1} & 0.005            & \textbf{1.887} \\
v2 &      623 & 1{,}742 &  2{,}360 & \textbf{5} & \textbf{0.803}   & 0.287          \\
\bottomrule
\end{tabular}\\[8pt]

\textit{Table 3b. Set-similarity indices.}\\[2pt]
\begin{tabular}{@{}l r r@{}}
\toprule
Window & Jaccard & Dice \\
\midrule
v1 & 0.000049 & 0.000099 \\
v2 & 0.002119 & 0.004228 \\
\bottomrule
\end{tabular}
\end{table}

Window periods: v1 = 11-25 Jun 2026 (strict cohort observation window;
the rejection stream covers only 24-25 Jun); v2 = 29 Jun-16 Jul 2026
(strict cohort observation window, fully covered by the rejection
stream). The two windows are drawn from non-overlapping time periods and
by two different detector versions.

Figure 1 visualises Table 3. The left panel reports absolute counts on a
log scale; the right panel reports the observed overlap proportions.
Both panels use a colour-plus-hatch pattern encoding for colour-blind
interpretability.

\begin{figure}[htbp]
\centering
\includegraphics[width=\linewidth]{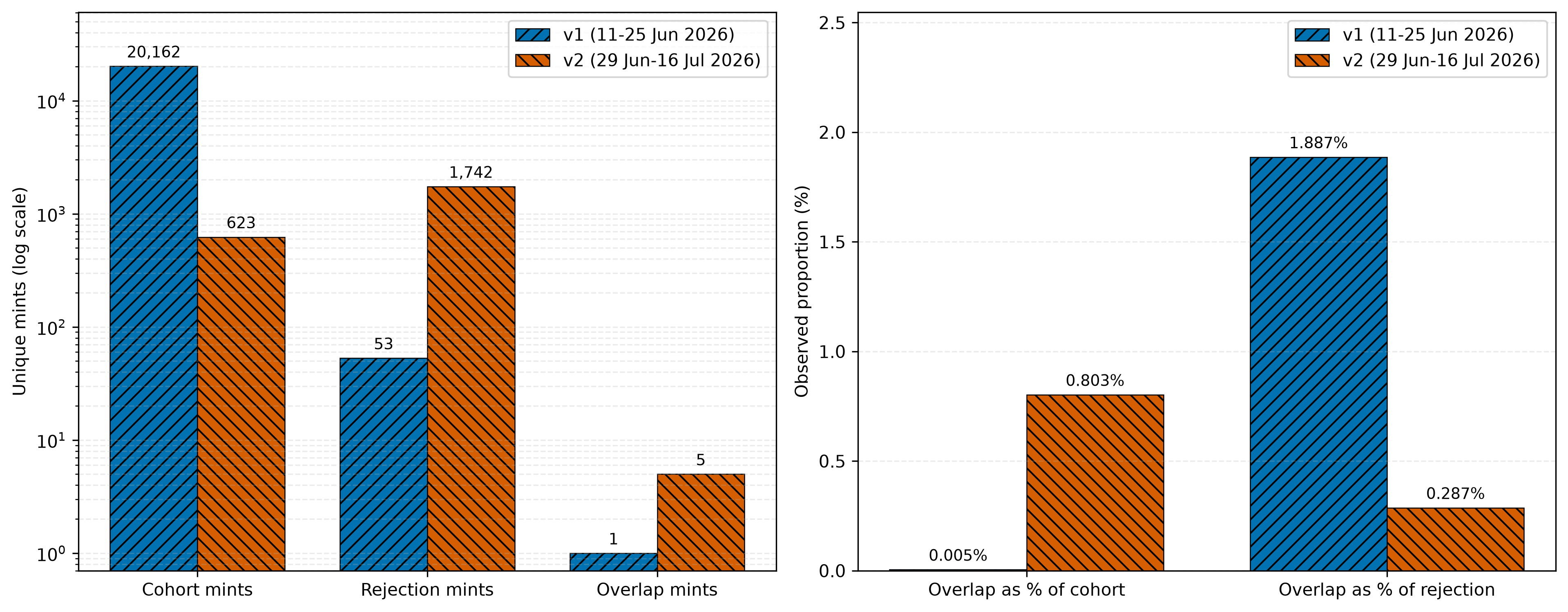}
\caption{Two-window descriptive analysis of cross-pipeline observation coverage on the Solana pump.fun launchpad. Left panel: absolute counts of cohort mints, rejection mints, and overlap mints on each of the two windows (log scale). Right panel: overlap counts expressed as observed proportions of the cohort and rejection sets. Windows: v1 = 11-25 Jun 2026 (blue, forward-hatch); v2 = 29 Jun-16 Jul 2026 (vermillion, back-hatch). No confidence intervals are shown; the observation-selection process on each side is deterministic (Section 4.3). In v1 the rejection stream covers only the final 14.25 hours of the cohort window, so the v1 cohort-side bar in the right panel reflects that gap and is not comparable with v2 (Section 5.3).}
\end{figure}

\subsection{v1↔v2 cohort intersection}\label{v1v2-cohort-intersection}

The intersection between the v1 and v2 cohort mint sets is \textbf{0}.
This is expected: the v2 window starts 3d 20h after the v1 window ends,
and cohort detection is a per-launch signature that resets across time
windows. The zero overlap does not speak to whether the underlying
wallet cohorts persist across windows, only to whether the same tokens
were touched.

\subsection{Rejection-stream coverage of the cohort
windows}\label{rejection-stream-coverage-of-the-cohort-windows}

The strict cohort window is set by the cohort detections alone. Table 4
reports how much of each window the rejection stream actually covers.
The common-coverage interval is the part of the strict cohort window
during which the rejection stream was recording: it runs from the later
of the window start and the first rejection event in the file to the
earlier of the window end and the last rejection event in the file.
Coverage is the length of that interval as a share of the window. Table
4 also reports the longest gap between successive rejection events
inside the interval, and the overlap after both mint sets are restricted
to the interval (cohort mints by \texttt{detected\_at}, rejection mints
by \texttt{rejectTs}; Supplementary Table S5).

\begin{table}[htbp]
\caption{Rejection-stream coverage of the strict cohort windows, and overlap within the common-coverage interval (Spec A; observed counts, no confidence intervals).}
\centering\small
\begin{tabular}{@{}l r r r r r r r@{}}
\toprule
Window & Hours & Covered (h) & Coverage (\%) & Max gap (h) & Cohort$^{a}$ & Rejection$^{a}$ & Overlap$^{a}$ \\
\midrule
v1 & 321.03 & 14.25 & 4.44 & 0.58 & 1{,}113 & 53 & 0 \\
v2 & 420.81 & 420.81 & 100.00 & 2.16 & 623 & 1{,}742 & 5 \\
\bottomrule
\end{tabular}
\par\smallskip\footnotesize $^{a}$ Mints within the common-coverage interval. Hours: length of the strict cohort window. Covered: length of the common-coverage interval. Coverage: covered hours as a percentage of the window. Max gap: longest interval between successive rejection events inside the common-coverage interval.
\end{table}

In v2 the rejection stream records throughout the cohort window; the
longest gap between successive rejection events is 2.2 hours. In v1 it
records only from 24 June 16:34 UTC, the end of the collection gap
described in Section 3.2, so it covers the final 14.25 hours (4.4\%) of
the 321-hour cohort window. Most v1 cohort mints were therefore detected
while the rejection pipeline was not recording, and the v1 cohort-side
proportion (1 of 20,162, or 0.005\%) mainly reflects the gap. It is not
a measure of cross-pipeline coverage, and the magnitude comparisons in
Section 6 use the rejection-side proportions instead. The v1
rejection-side proportion (1 of 53) is less affected, because all 53
rejection mints were observed while the cohort pipeline was running.

Restricting both mint sets to the v1 common-coverage interval (24 June
16:34 UTC to 25 June 06:49 UTC, 14.25 hours) leaves 1,113 cohort mints
and the same 53 rejection mints, with no overlap. This count is
uninformative. The restriction drops the 19,049 v1 cohort mints detected
before the interval, including the single v1 overlap mint
(\texttt{5qjZ...pump}), which the cohort pipeline detected on 23 June
and the rejection stream first rejected 25.9 hours later; in v2 the
first rejection of the five overlap mints followed their cohort
detection by between \(-1.8\) and 91.9 hours (median 2.3 hours;
Supplementary Table S2). A short interval therefore biases the overlap
toward zero. In addition, at the v2 rejection-side proportion (0.287\%)
the expected overlap among 53 rejection mints would be about 0.15 mints,
so an observed count of zero neither supports nor contradicts the v2
result.

\subsection{Sensitivity to the timestamp
gate}\label{sensitivity-to-the-timestamp-gate}

On v1, all four strict-window rules give 53 rejection mints and 1
overlap (1.887\% of rejection), and all four wider-window rules and the
file-level count give 83 rejection mints and 2 overlaps (2.410\%;
Supplementary Table S3). The additional overlapping mint recovered under
the wider window is \texttt{AvxF...pump}, whose earliest rejection
timestamp is 2026-06-25T19:49:28Z, 13h 00m 03s past the upper edge of
the strict cohort window. The strict and wider values differ only in the
numerator; both are single-digit intersections on small denominators. On
v2 the strict and file-level counts differ by at most 14 mints
(1,742-1,749 vs 1,756), with the overlap fixed at 5.

\section{Discussion}\label{discussion}

\subsection{What the finding is}\label{what-the-finding-is}

The two pipeline outputs touch \emph{nearly}, not entirely, disjoint
mint populations. Under the primary specification the overlap is a
single-digit count in both windows (1 mint in v1, 5 in v2):
approximately 0.8\% of the cohort side and 0.3\% of the rejection side
in v2, and one mint out of 53 on the v1 rejection side. Both windows
agree on \emph{direction}: the two pipelines share tokens in the
single-percent range or below, not in the tens-of-percent range. Because
the v1 rejection stream covers only 4.4\% of its cohort window (Section
5.3), the v1 window supports the direction of the finding but carries
much less weight than v2.

The agreement is a within-programme, direction-level statement, not an
independent replication and not an estimate of a population rate. The
two pipelines share collector infrastructure, licence and author
(Section 3.2), and the observed magnitudes differ (rejection-side
1.887\% in v1 vs 0.287\% in v2; the v1 cohort-side value is not
comparable, Section 5.3). The overlap is small but not zero, and no
inferential claim about an underlying overlap rate is made (Section
4.3). Treating consistent-direction findings from two runs of one
programme as if they came from independent teams would ignore the
non-standard errors documented by Menkveld et al.~{[}36{]}.

\subsection{Why the two pipeline outputs are nearly
disjoint}\label{why-the-two-pipeline-outputs-are-nearly-disjoint}

Four factors offer plausible descriptive explanations for the low
overlap:

\begin{enumerate}
\def\labelenumi{\arabic{enumi}.}
\tightlist
\item
  \textbf{Lifecycle stage.} Cohort detection collects pre-graduation
  buyer-side signatures; rejection filtering is dominated by
  post-graduation tradability-side signatures (Section 3.5 discloses a
  small pre-graduation residual on the rejection side: 5,040 rows / 41
  mints with \texttt{dexId=pumpfun}). A mint that never graduates and
  never reaches a graduated DEX venue is unlikely to enter the rejection
  stream at meaningful scale. Marino et al.~{[}1{]} model which pump.fun
  tokens graduate; the present paper observes the lifecycle only through
  the two pipelines.
\item
  \textbf{Filter surface.} The rejection filter set (price-trend,
  market-capitalisation, time-of-day and order-flow filters; Section
  3.2) rules out a large fraction of the graduated-mint universe on
  grounds unrelated to cohort behaviour. Cohorts, being pre-graduation
  signatures, are approximately orthogonal to these tradability filters.
\item
  \textbf{Set-size asymmetry.} The two sides differ greatly in size, in
  opposite directions in the two windows (v1: 20,162 cohort vs 53
  strict-window rejection mints; v2: 623 vs 1,742). The asymmetry bounds
  the overlap proportion achievable on each side; it is not a mechanism
  for the low overlap.
\item
  \textbf{Coverage (v1 only).} In v1 the rejection pipeline was
  recording for only the final 14.25 hours of the 321-hour cohort window
  (Section 5.3), so most v1 cohort mints could not have entered the v1
  rejection file.
\end{enumerate}

Apart from coverage, which Section 5.3 measures, none of these
mechanisms is measured directly. The overlap measurement is agnostic to
the mechanism.

\subsection{Implications for on-chain
measurement}\label{implications-for-on-chain-measurement}

The direct consequence is an accounting statement. On v2, an analyst who
used only the cohort pipeline would have observed 623 mints; only the
rejection pipeline, 1,742 mints; both, 2,360 unique mints. Combining the
pipelines expands the observed set by approximately \textbf{35\%}
relative to the larger rejection stream (2,360 / 1,742 = 1.355) and by
approximately \textbf{3.8×} relative to the smaller cohort stream (2,360
/ 623 = 3.79). This union is the set of mints visible to the two
pipelines, not the pump.fun launch universe, which is not measured here;
a dual-frame reweighting of the union {[}33{]} would require a
source-inclusion model that this paper does not attempt.

Three implications follow for studies that build their own on-chain
collectors, as the studies reviewed in Section 2 do. First, a token set
produced by one collector describes that collector's configuration as
much as the market; findings about ``memecoins'' or ``launches'' should
name the collector and its filters. Second, a collector's own coverage
of the analysis window should be reported, because a single collection
gap can make a proportion uninformative, as Table 4 shows for v1. Third,
when more than one collector is available, the overlap and union of
their outputs under a stated timestamp gate are cheap to compute and
should be reported before whole-market statements are made. None of
these implications depends on the specific magnitudes reported here.

\section{Limitations and corrections}\label{limitations-and-corrections}

\subsection{Limitations}\label{limitations}

\begin{itemize}
\tightlist
\item
  \textbf{Launch universe not measured.} The paper measures the union
  and intersection of two pipeline outputs. It does not measure the
  completeness or recall of either pipeline against the full pump.fun
  launch universe; that would require a whole-launchpad collection of
  the kind used in large-scale studies {[}2{]}.
\item
  \textbf{Pipelines not statistically independent.} The two pipelines
  share collector infrastructure, licence and author, and the rejection
  files are extracts of the public RED-REJECT-2026-v2 dataset, whose
  collection also underlies a companion preprint's data (Section 3). The
  two-window agreement is descriptive, not a statistical or independent
  replication.
\item
  \textbf{No confidence intervals.} The observation-selection process on
  each side is deterministic and admits no defensible sampling model
  (Section 4.3); the proportions are observed proportions of observed
  sets, and no significance is claimed.
\item
  \textbf{Partial coverage and small n in v1.} The v1 rejection stream
  covers only the final 14.25 hours (4.4\%) of the v1 cohort window and
  contains 53 mints with one overlap. The v1 window contributes a
  direction-of-finding observation only; the substantive comparison
  rests on v2.
\item
  \textbf{\texttt{dexId=pumpfun} residual.} 5,040 rejection rows / 41
  mints on v2 are pre- or during-graduation observations (Section 3.5),
  which tempers the characterisation of the rejection stream as
  post-graduation.
\item
  \textbf{Configuration dependence.} The magnitudes depend on the
  specific detector version, subscription set and filter thresholds
  documented in Section 3; other collector configurations would
  generically give different overlaps on the same event stream.
\end{itemize}

\subsection{Corrections to the earlier
preprint}\label{corrections-to-the-earlier-preprint}

This version corrects the earlier preprint (arXiv:2609.18975 version 1,
SSRN 7128818 and Zenodo record 10.5281/zenodo.21399919). The v1 result
is corrected from 2 overlapping mints among 83 rejection mints to 1
among 53: the earlier figures used the wider extraction interval of
Section 4.4 rather than the strict cohort window. The v2 rejection count
is 1,742 mints (Spec A), not 1,743; the v2 overlap is unchanged. The v1
coverage gap (Section 5.3) was not disclosed earlier. A Poisson-null
significance test and two further analyses are withdrawn, and several
statements of the earlier abstract are withdrawn: that the pipelines are
``independent'', that overlap ``remains under 1 percent'' as a threshold
statement, that the separation is ``structural rather than
statistical'', and that a ``time-window artifact'' and three other
explanations were ``systematically ruled out''. Supplementary Section
S7, gives the full list with reasons.

\section{Conclusion}\label{conclusion}

Two separately configured on-chain observation pipelines within one
research programme, a cohort-detection pipeline and a
rejection-filtering pipeline, were compared on the Solana pump.fun
launchpad over two consecutive windows. Under a strict cohort-window
definition their output mint sets are nearly disjoint: 5 overlapping
mints among 623 cohort and 1,742 rejection mints in the fully covered
June-July 2026 window, and 1 overlapping mint among 20,162 cohort and 53
rejection mints in the June 2026 window, where the rejection stream
covered only the final 14.25 hours of the cohort window. The agreement
between the windows is descriptive, not a statistical replication.
Combining the two pipelines in the fully covered window expands the
observed set by about 35\% relative to the larger pipeline and 3.8×
relative to the smaller.

The implication for on-chain research is limited but concrete: an
analyst who characterises a permissionless launchpad from a single
collector should treat that collector's token set as one
configuration-dependent slice, report the collector's coverage of the
analysis window, and compare collectors where possible before making
whole-market statements. The paper does not claim independence between
the pipelines, does not measure the pump.fun launch universe and does
not model the observation-selection mechanism.

\section*{Declarations}\label{declarations}
\addcontentsline{toc}{section}{Declarations}

\textbf{CRediT authorship contribution statement.} Arati Uday Kamat:
Conceptualization, Methodology, Software, Formal analysis, Data
curation, Visualization, Writing -- original draft, Writing -- review \&
editing.

\textbf{Funding.} This research did not receive any specific grant from
funding agencies in the public, commercial, or not-for-profit sectors.

\textbf{Declaration of competing interest.} The author designed and
operates the paper-trading system that generated the rejection events;
no live-money trades were executed against the tokens in the rejection
data. The author has pending provisional patent applications potentially
related to this line of research. The author has no financial
relationship with the pump.fun protocol, with Solana Labs, with any
decentralised exchange named in the manuscript, or with any market
participant identified in the data.

\textbf{Data availability.} The data and code are openly available in
the reproducibility package \texttt{cross\_pipeline\_coverage\_v2/}
{[}7{]}, deposited at Zenodo as version 2 of the record with concept DOI
\href{https://doi.org/10.5281/zenodo.21399918}{10.5281/zenodo.21399918};
the concept DOI resolves to the latest version. The package contains the
four input files (pinned by SHA-256; Supplementary Table S6), the
reconstruction script \texttt{reconstruct\_overlap.py} and the
containment check \texttt{verify\_red\_reject\_containment.py} (Python
3.12, standard library only; MIT License), the outputs, and a data
dictionary. The containment check additionally uses the public
RED-REJECT-2026-v2 file. The v1 cohort file is a member of the public
RED-COHORT-2026-v1 dataset (concept DOI
\href{https://doi.org/10.5281/zenodo.20978741}{10.5281/zenodo.20978741}),
and the rejection files are extracts of the public RED-REJECT-2026-v2
dataset (concept DOI
\href{https://doi.org/10.5281/zenodo.21402476}{10.5281/zenodo.21402476}).
Data are released under Creative Commons Attribution 4.0 International
(CC BY 4.0).

\textbf{Ethics statement.} The study analyses public on-chain data
(Solana mint addresses and associated observation events). No human
participants were involved and no personally identifiable information
was collected.

\textbf{Use of generative AI.} Portions of the manuscript's prose were
prepared with authoring assistance from a large language model; the
author verified every claim, checked every number against the
reproduction script, and takes full responsibility for the content.

\textbf{Supplementary material.} The supplementary material
(\texttt{ESM\_1.pdf}) contains: data-quality diagnostics on the four
input files (Table S1), the addresses and detection and rejection times
of the overlap mints (Table S2), the full timestamp-rule sensitivity
table for both windows (Table S3), the venue breakdown of the v2
rejection stream (Table S4), the rejection-stream coverage of both
windows (Table S5), an inventory of the data release with input
checksums (Table S6), the analyses withdrawn from or corrected in the
earlier preprint (Section S7), and reproduction instructions (Section
S8).

\section*{References}\label{references}
\addcontentsline{toc}{section}{References}

{[}1{]} Marino, G., Naviglio, M., Tarantelli, F., \& Lillo, F. (2026).
\emph{Predicting the success of new crypto-tokens: The Pump.fun case}
(arXiv:2602.14860) {[}Preprint{]}. arXiv.
\url{https://doi.org/10.48550/arXiv.2602.14860}

{[}2{]} Szwajcok, N., Tsuchiya, T., Liu, E., Soska, K., Payer, M., \&
Christin, N. (2026). \emph{Meme coin factories: Uncovering large-scale
manipulations on pump.fun} (arXiv:2609.10246) {[}Preprint{]}. arXiv.
\url{https://doi.org/10.48550/arXiv.2609.10246}

{[}3{]} Srifa, S., Yanovich, Y., Vasilyev, R., Rupasinghe, T., \&
Amelin, V. (2025). Rug pull detection on decentralized exchange using
transaction data. \emph{Blockchain: Research and Applications},
\emph{6}(3), Article 100275.
\url{https://doi.org/10.1016/j.bcra.2025.100275}

{[}4{]} Kalacheva, A., Kuznetsov, P., Vodolazov, I., \& Yanovich, Y.
(2026). Detecting rug pulls in decentralized exchanges: The rise of meme
coins. \emph{Blockchain: Research and Applications}, \emph{7}(2),
Article 100336. \url{https://doi.org/10.1016/j.bcra.2025.100336}

{[}5{]} Cernera, F., La Morgia, M., Mei, A., \& Sassi, F. (2022).
\emph{Token spammers, rug pulls, and sniperbots: An analysis of the
ecosystem of tokens in Ethereum and in the Binance Smart Chain (BNB)}
(arXiv:2206.08202) {[}Preprint{]}. arXiv.
\url{https://doi.org/10.48550/arXiv.2206.08202}

{[}6{]} Vidal-Tomás, D. (2022). Which cryptocurrency data sources should
scholars use? \emph{International Review of Financial Analysis},
\emph{81}, Article 102061.
\url{https://doi.org/10.1016/j.irfa.2022.102061}

{[}7{]} Kamat, A. U. (2026d). \emph{Cross-pipeline coverage on the
Solana pump.fun launchpad: Reproducibility package} (Version 2.0.0)
{[}Data set{]}. Zenodo. \url{https://doi.org/10.5281/zenodo.21399918}

{[}8{]} Daian, P., Goldfeder, S., Kell, T., Li, Y., Zhao, X., Bentov,
I., Breidenbach, L., \& Juels, A. (2020). Flash Boys 2.0: Frontrunning
in decentralized exchanges, miner extractable value, and consensus
instability. In \emph{2020 IEEE Symposium on Security and Privacy (SP)}
(pp.~910-927). IEEE. \url{https://doi.org/10.1109/SP40000.2020.00040}

{[}9{]} Qin, K., Zhou, L., \& Gervais, A. (2022). Quantifying blockchain
extractable value: How dark is the forest? In \emph{2022 IEEE Symposium
on Security and Privacy (SP)} (pp.~198-214). IEEE.
\url{https://doi.org/10.1109/SP46214.2022.9833734}

{[}10{]} Werner, S., Perez, D., Gudgeon, L., Klages-Mundt, A., Harz, D.,
\& Knottenbelt, W. (2022). SoK: Decentralized finance (DeFi). In
\emph{Proceedings of the 4th ACM Conference on Advances in Financial
Technologies (AFT '22)} (pp.~30-46). ACM.
\url{https://doi.org/10.1145/3558535.3559780}

{[}11{]} Zhou, L., Xiong, X., Ernstberger, J., Chaliasos, S., Wang, Z.,
Wang, Y., Qin, K., Wattenhofer, R., Song, D., \& Gervais, A. (2023).
SoK: Decentralized finance (DeFi) attacks. In \emph{2023 IEEE Symposium
on Security and Privacy (SP)} (pp.~2444-2461). IEEE.
\url{https://doi.org/10.1109/SP46215.2023.10179435}

{[}12{]} Öz, B., Torres, C. F., Schlegel, C., Mazorra, B., Gebele, J.,
Rezabek, F., \& Matthes, F. (2025). Cross-chain arbitrage: The next
frontier of MEV in decentralized finance. \emph{Proceedings of the ACM
on Measurement and Analysis of Computing Systems}, \emph{9}(3), 1-33.
\url{https://doi.org/10.1145/3771566}

{[}13{]} Lehar, A., \& Parlour, C. A. (2025). Decentralized exchange:
The Uniswap automated market maker. \emph{The Journal of Finance},
\emph{80}(1), 321-374. \url{https://doi.org/10.1111/jofi.13405}

{[}14{]} Angeris, G., \& Chitra, T. (2020). Improved price oracles:
Constant function market makers. In \emph{Proceedings of the 2nd ACM
Conference on Advances in Financial Technologies (AFT '20)} (pp.~80-91).
ACM. \url{https://doi.org/10.1145/3419614.3423251}

{[}15{]} Angeris, G., Evans, A., Chitra, T., \& Boyd, S. (2022). Optimal
routing for constant function market makers. In \emph{Proceedings of the
23rd ACM Conference on Economics and Computation (EC '22)}
(pp.~115-128). ACM. \url{https://doi.org/10.1145/3490486.3538336}

{[}16{]} Heimbach, L., \& Wattenhofer, R. (2022). SoK: Preventing
transaction reordering manipulations in decentralized finance. In
\emph{Proceedings of the 4th ACM Conference on Advances in Financial
Technologies (AFT '22)} (pp.~47-60). ACM.
\url{https://doi.org/10.1145/3558535.3559784}

{[}17{]} Cong, L. W., Li, Y., \& Wang, N. (2021). Tokenomics: Dynamic
adoption and valuation. \emph{The Review of Financial Studies},
\emph{34}(3), 1105-1155. \url{https://doi.org/10.1093/rfs/hhaa089}

{[}18{]} Cong, L. W., Li, Y., \& Wang, N. (2022). Token-based platform
finance. \emph{Journal of Financial Economics}, \emph{144}(3), 972-991.
\url{https://doi.org/10.1016/j.jfineco.2021.10.002}

{[}19{]} Anton, E., Aptyka, M., Oesterreich, T. D., \& Teuteberg, F.
(2024). To the moon with Dogecoin! Disentangling the causalities behind
extrinsic and intrinsic motivations for memecoin investments.
\emph{Journal of Decision Systems}, 1-35.
\url{https://doi.org/10.1080/12460125.2024.2348936}

{[}20{]} Conlon, T., \& Corbet, S. (2025). Memecoin contagion:
Irrationality, illicit behaviour, and cryptocurrency risk. \emph{Finance
Research Letters}, \emph{86}, Article 108264.
\url{https://doi.org/10.1016/j.frl.2025.108264}

{[}21{]} Dhawan, A., \& Putniņš, T. J. (2023). A new wolf in town?
Pump-and-dump manipulation in cryptocurrency markets. \emph{Review of
Finance}, \emph{27}(3), 935-975.
\url{https://doi.org/10.1093/rof/rfac051}

{[}22{]} Tsuchiya, T. (2021). Profitability of cryptocurrency pump and
dump schemes. \emph{Digital Finance}, \emph{3}(2), 149-167.
\url{https://doi.org/10.1007/s42521-021-00034-6}

{[}23{]} Cordoba Otalora, F. (2025). Student research abstract:
Subculture-driven speculation: Meme pump dot fun metric on the Solana
blockchain. In \emph{Proceedings of the 40th ACM/SIGAPP Symposium on
Applied Computing (SAC '25)} (pp.~391-395). ACM.
\url{https://doi.org/10.1145/3672608.3707997}

{[}24{]} Ali, Z. (2025). \emph{Adverse selection and market failure in
Solana token markets: An empirical and behavioral analysis} {[}SSRN
Working Paper No.~5239105{]}. SSRN.
\url{https://doi.org/10.2139/ssrn.5239105}

{[}25{]} Meiklejohn, S., Pomarole, M., Jordan, G., Levchenko, K., McCoy,
D., Voelker, G. M., \& Savage, S. (2013). A fistful of bitcoins:
Characterizing payments among men with no names. In \emph{Proceedings of
the 2013 Internet Measurement Conference (IMC '13)} (pp.~127-140). ACM.
\url{https://doi.org/10.1145/2504730.2504747}

{[}26{]} Meiklejohn, S., Pomarole, M., Jordan, G., Levchenko, K., McCoy,
D., Voelker, G. M., \& Savage, S. (2016). A fistful of Bitcoins:
Characterizing payments among men with no names. \emph{Communications of
the ACM}, \emph{59}(4), 86-93. \url{https://doi.org/10.1145/2896384}

{[}27{]} Foley, S., Karlsen, J. R., \& Putniņš, T. J. (2019). Sex,
drugs, and Bitcoin: How much illegal activity is financed through
cryptocurrencies? \emph{The Review of Financial Studies}, \emph{32}(5),
1798-1853. \url{https://doi.org/10.1093/rfs/hhz015}

{[}28{]} Xia, P., Wang, H., Gao, B., Su, W., Yu, Z., Luo, X., Zhang, C.,
Xiao, X., \& Xu, G. (2021). Trade or trick? Detecting and characterizing
scam tokens on Uniswap decentralized exchange. \emph{Proceedings of the
ACM on Measurement and Analysis of Computing Systems}, \emph{5}(3),
1-26. \url{https://doi.org/10.1145/3491051}

{[}29{]} Lin, Z., Chen, J., Wu, J., Zhang, W., Wang, Y., \& Zheng, Z.
(2024). CRPWarner: Warning the risk of contract-related rug pull in DeFi
smart contracts. \emph{IEEE Transactions on Software Engineering},
\emph{50}(6), 1534-1547. \url{https://doi.org/10.1109/TSE.2024.3392451}

{[}30{]} Chao, A. (1987). Estimating the population size for
capture-recapture data with unequal catchability. \emph{Biometrics},
\emph{43}(4), 783-791. \url{https://doi.org/10.2307/2531532}

{[}31{]} Chao, A. (1989). Estimating population size for sparse data in
capture-recapture experiments. \emph{Biometrics}, \emph{45}(2), 427-438.
\url{https://doi.org/10.2307/2531487}

{[}32{]} Fellegi, I. P., \& Sunter, A. B. (1969). A theory for record
linkage. \emph{Journal of the American Statistical Association},
\emph{64}(328), 1183-1210.
\url{https://doi.org/10.1080/01621459.1969.10501049}

{[}33{]} Lohr, S. L., \& Rao, J. N. K. (2000). Inference from dual frame
surveys. \emph{Journal of the American Statistical Association},
\emph{95}(449), 271-280.
\url{https://doi.org/10.1080/01621459.2000.10473920}

{[}34{]} Westreich, D. (2012). Berkson's bias, selection bias, and
missing data. \emph{Epidemiology}, \emph{23}(1), 159-164.
\url{https://doi.org/10.1097/EDE.0b013e31823b6296}

{[}35{]} Peng, R. D. (2011). Reproducible research in computational
science. \emph{Science}, \emph{334}(6060), 1226-1227.
\url{https://doi.org/10.1126/science.1213847}

{[}36{]} Menkveld, A. J., Dreber, A., Holzmeister, F., Huber, J.,
Johannesson, M., Kirchler, M., Neusüß, S., Razen, M., Weitzel, U.,
Abad-Díaz, D., Abudy, M., Adrian, T., Aït-Sahalia, Y., Akmansoy, O.,
Alcock, J. T., Alexeev, V., Aloosh, A., Amato, L., Amaya, D., \ldots{}
Zwinkels, R. (2024). Nonstandard errors. \emph{The Journal of Finance},
\emph{79}(3), 2339-2390. \url{https://doi.org/10.1111/jofi.13337}

{[}37{]} Kamat, A. U. (2026b). \emph{RED-COHORT-2026-v1: A catalogue of
1,012 persistent wallet cohorts detected on the Solana Pump.fun
bonding-curve marketplace (June 11-25, 2026)} (Version 1.1.1) {[}Data
set{]}. Zenodo. \url{https://doi.org/10.5281/zenodo.20978741}

{[}38{]} Kamat, A. U. (2026a). \emph{Coordinated sniper cohorts on
Pump.fun: Detection of 1,012 persistent wallet rings and a
contamination-adjusted estimate of coordination-specific first-hour
buyer-flow lift} (arXiv:2607.02795) {[}Preprint{]}. arXiv.
\url{https://doi.org/10.48550/arXiv.2607.02795}

{[}39{]} Kamat, A. U. (2026c). \emph{RED-REJECT-2026-v2 (corrected
release): A 96-day public corpus of algorithmic filter rejections with
post-rejection follow-up samples on the Solana pump.fun ecosystem}
(Version v2.2) {[}Data set{]}. Zenodo.
\url{https://doi.org/10.5281/zenodo.21402476}

\end{document}